# Dynamics of macrospin under periodic field and spin transfer torque

Weiyi Ren and Hao Yu[*]

Department of Mathematical Sciences, Xi'an Jiaotong-Liverpool University, Suzhou, China, 215123

*Email address: hao.yu@xjtlu.edu.cn

**Abstract**

The dynamics of a macrospin model for a single magnetic domain is investigated in two cases: (i) under the action of a periodic magnetic field and (ii) under the external field plus a spin transfer torque from spin-polarized current or spin current. It has been proved that (i) without spin transfer torque, the trajectory of magnetization (hysteresis) is always a closed curve oscillating between two stable points and following the same path each time; (ii) under the action of a constant field (or with small periodic perturbation) and a constant spin transfer torque, the spin would always turn to the stable solution finally, which is independent of the parameters or initial conditions; (iii) considering a periodic oscillating field plus a spin transfer torque, the system can also have a stable solution around a limit circle, but on some special points on the attractor, the system may be sensitive to the initial conditions and have no limit circle.

## 1. Introduction

To understand the dynamics of magnetization of nanomagnets is significant to spintronic devices, especially for the high density and high speed spin-based data storage or logic applications. The dynamic response of a nanomagnet to oscillating magnetic field or spin polarized current is quite different from that of static-equilibrium condition in terms of magnetization.

The Landau-Lifshitz-Gilbert equation [1, 2] is a nonlinear differential equation describing the evolution of magnetization. The magnetization vector is processing about the axis of effective magnetic field and the Gilbert damping term provides a torque to make the magnetization moment align with the effective field. The effective filed can be determined by the variation of free energy with respect to the magnetization. The free energy consists of the short-range exchange, anisotropy and demagnetization energy[3,4]. By introducing extra torque term, e.g., spin transfer torque [5], the equation can describe the magnetization driven by a spin-polarized current or spin current.

In this article, we will study the dynamic magnetization behaviors of a macrospin, usually corresponding to a single domain, under a periodic field with and without the action of spin transfer toque.

## 2. Dynamics of single macrospin under periodic field

Consider the Landau-Lifshitz-Gilbert Equation of the magnetization vector $M$:

$$\frac{d\mathbf{M}}{dt} = -\gamma_0(\mathbf{M} \times \mathbf{H}) + \lambda\gamma_0\left(\mathbf{M} \times \frac{d\mathbf{M}}{dt}\right) \quad (1)$$

where $\gamma_0 = \mu_0 g \mu_B/\hbar$, $\lambda \ll 1$ is the damping coefficient.

By computing $\mathbf{M} \cdot d\mathbf{M}/dt = 0$, we could simply get that $|\mathbf{M}|$=constant. Defining the unit magnetization vector (macrospin of a single domain) $\mathbf{S} = \mathbf{M}/|\mathbf{M}| = (S_1, S_2, S_3)$ and by substitution for $d\mathbf{S}/dt$ it turns out to be:

$$(1 + \lambda^2\gamma_0^2)\frac{d\mathbf{S}}{dt} = -\gamma_0(\mathbf{S} \times \mathbf{H}) - \lambda\gamma_0^2 \cdot \mathbf{S} \times (\mathbf{S} \times \mathbf{H}) \quad (2)$$

After rescaling the time: let $\tilde{t} = \frac{t\gamma_0}{1+\lambda^2\gamma_0^2}$ (To simplify, we still use $t$ instead of $\tilde{t}$), it can be written as:

$$\frac{d\mathbf{S}}{dt} = -\mathbf{S} \times \mathbf{H} - \lambda\gamma_0 \cdot \mathbf{S} \times (\mathbf{S} \times \mathbf{H}) \quad (3)$$

Suppose the external field is applied in the $z$-axis: $H = (0, 0, f(t))$, where $f(t)$ is a continuous periodic function with period $T$, let $\lambda\gamma_0 = \beta$, $z = S_3$. Along the $z$-direction, it could separate the variable to get:

$$\frac{dz}{dt} = \beta f(t)(1 - z^2)$$

Suppose $\int_0^T f(t)\,dt = a > 0$, it could be derived by integration that $\lim_{t\to\infty} z = 1$. For the same reason, if $\int_0^T f(t)\,dt = b < 0$, then $\lim_{t\to\infty} z = -1$. This result indicates that no matter what initial condition takes, $S$ would always turn to $(0,0,\pm 1)$ after long enough time.

The situation becomes complicated when $\int_0^T f(t)\,dt = 0$. Also by integration:

$$\int_{z_{t_k}}^{z_{t_k+T}} \frac{1}{1-z^2}\,dz = \int_{t_k}^{t_k+T} \beta f(t)\,dt = 0 \Rightarrow z_{t_k} = z_{t_k+T} \tag{4}$$

Which indicates that $z(t) = S_3(t)$ is a periodic function.

Then compute $dS_1/dt$ and $dS_2/dt$ separately and eliminate $f(t)$:

$$S_2 \frac{dS_1}{dt} - S_1 \frac{dS_2}{dt} = -f(t)(S_1^2 + S_2^2) = -\frac{1}{\beta}\frac{dS_3}{dt} \tag{5}$$

Using $S_3 = \pm\sqrt{1 - S_2^2 - S_1^2}$, we can calculate $dS_3/dS_1$ and $dS_3/dS_2$ which turns out to be

$$\mp \frac{S_2\sqrt{1 - S_2^2 - S_1^2}}{S_1} \pm \frac{S_1\sqrt{1 - S_2^2 - S_1^2}}{S_2} = -\frac{1}{\beta}$$

Pick up arbitrary time $t$ in the first period $T$, then $S_3(t) = S_3(t + kT)$, and $(S_1(t), S_2(t))$ and $(S_1(t + kT), S_2(t + kT))$ are two solutions of the following function set:

$$\begin{cases} S_1^2 + S_2^2 = a \\ \dfrac{S_1}{S_2} - \dfrac{S_2}{S_1} = b \end{cases}$$

Where $a, b$ are constant. The function set has at most four solutions. Suppose $(x_0, y_0)$ is a solution, then the other three solutions are $(-x_0, -y_0), (-y_0, x_0), (y_0, -x_0)$. It indicates that, for the same $S_3$, $(S_1, S_2)$ could take only four different values. Finally, we use the property of continuous function to show that the system only has periodic unique solution.

Suppose in the first period, $S_3$ gets it's maximum at $t = t_0$, and $S_3 \neq 1$ for all $t$ (unless $f(t) = \infty$). By the symmetry of four possible solutions of $(S_1, S_2)$ at time $t$, we can get the minimum distance between them is $\sqrt{2}(S_1^2 + S_2^2) \geq \sqrt{2}(1 - S_3(t_0)^2 > 0$, so the four possible trajectory of $(S_1, S_2)$ never intersect. Because $S$ is a continuous function, so $(S_1, S_2)$ could only pick up one of the solution trajectory and oscillate.

In conclusion, when $\int_0^T f(t)\,dt = 0$, the trajectory is a curve between the initial point and the point when $S_3$ takes the maximum. It oscillates between the two points and follows the same path each time.

## 3. Dynamics of single macrospin with spin transfer torque

By considering the spin transfer torque[6,7], the Landau-Lifshitz equation turns into:

$$\frac{d\boldsymbol{S}}{dt} = \boldsymbol{S} \times (\boldsymbol{H} - \lambda \boldsymbol{H} \times \boldsymbol{S} + \alpha \boldsymbol{S} \times \boldsymbol{S}_p) \tag{6}$$

Where $\boldsymbol{S}_p$ is the pinned direction of the spin-polarized current and $\alpha > 0$ is the constant parameter corresponding to the strength of the current, and $\lambda > 0$ is the damping parameter.

In a similar way, by computing $\boldsymbol{S} \cdot d\boldsymbol{S}/dt$ we can see that $|\boldsymbol{S}|$ is a constant. Then we can assume $|\boldsymbol{S}| = 1$. For the symmetry of the sphere, we can apply the external field in the direction of $z$-axis, and set $\boldsymbol{S}_p$ in the $y$-$z$ plane which makes an angle $\psi$ with the $z$-axis.

Suppose that the external field is a constant with $\boldsymbol{H} = (0,0,h_z)$, $\boldsymbol{S}_p = (S_{px}, 0, S_{pz})$ and $\boldsymbol{S} = (S_1, S_2, S_3)$. Because it's an autonomous sytem of ODEs on a sphere, by the Poincare-Bendixson theorem, the system contains only equilibrium, periodic solution or saddle cycle.

If the spin-polarized current is relatively small, then the external field predominate the process. The spin would do spiral motion and converge to the limit point finally. If the external field is relatively small, then the spin would converge directly to the limit point by the effect of spin transfer torque. The first term $\boldsymbol{S} \times \boldsymbol{H}$ would make the spin spinning around the $z$-axis. The remaining two torques, namely the damping torque and the spin transfer torque, would attract the spin to the direction of $z$-axis or the direction of the current, respectively.

By computing the equilibrium points: $(\frac{dS_1}{dt}, \frac{dS_2}{dt}, \frac{dS_3}{dt}) = (0,0,0)$, we have four solutions for this system, and $S_2$ could be expressed as:

$$S_2 = \frac{h_z^2 \lambda^2 \pm \sqrt{\left(h_z^2\lambda^2 + h_z^2 + 2\alpha h_z \lambda S p_z + \alpha^2 S_{px}^2 + \alpha^2 S_{pz}^2\right)^2 - 4\alpha^2 h_z^2 S_{px}^2} + h_z^2 + 2\alpha h_z \lambda S_{pz} + \alpha^2 S_{px}^2 + \alpha^2 S_{pz}^2}{2\alpha h_z S_{px}}$$

If $|S_2| \leq 1$, then the equilibriums exist. If the system does not have any equilibrium, then by Poincare-Bendixson theorem, there must exists a limit cycle.

$$\begin{aligned}\Delta &= \left(h_z^2\lambda^2 + h_z^2 + 2\alpha h_z\lambda Sp_z + \alpha^2 S_{px}^2 + \alpha^2 S_{pz}^2\right)^2 - 4\alpha^2 h_z^2 S_{px}^2 \\ &= \left((h_z + \alpha S_{pz}^2)^2 + h_z^2\lambda^2 + 2\alpha h_z\lambda Sp_z + \alpha^2 S_{px}^2\right)\left((h_z - \alpha S_{pz}^2)^2 + h_z^2\lambda^2 + 2\alpha h_z\lambda Sp_z + \alpha^2 S_{px}^2\right) > 0\end{aligned}$$

To simplify the equation, we set $h_z^2\lambda^2 + h_z^2 + 2\alpha h_z\lambda Sp_z + \alpha^2 S_{px}^2 + \alpha^2 S_{pz}^2 = A$ and $2\alpha h_z S_{px} = B$, then

$$S_2 = \frac{A \pm \sqrt{A^2 - B^2}}{B} = \frac{A}{B} \pm \sqrt{\frac{A^2}{B^2} - 1}$$

Also by $\Delta > 0$ we can get that $A^2 > B^2 \Rightarrow \left|\frac{A}{B}\right| > 1$, by multiplying two solution for $S_2$, we could get:

$$\left(\frac{A}{B} + \sqrt{\frac{A^2}{B^2} - 1}\right) \cdot \left(\frac{A}{B} - \sqrt{\frac{A^2}{B^2} - 1}\right) = 1$$

Which shows that there is always only one solution of $S_2$ satisfying $|S_2| < 1$, so the system could only have at most two symmetry equilibriums: $(S_1, S_2, S_3), (-S_1, S_2, -S_3)$. Suppose the system do not have any equilibriums and assume that $S_3^2 + S_2^2 > 1$, then $S_1$ is an imaginary number, and $\alpha S_1 S_3 S_{px} - \alpha S_1^2 S_{pz} - \alpha S_2^2 S_{pz} - \lambda h_z S_1^2 - \lambda h_z S_2^2 = 0$ indicates that $\alpha S_3 S_{px} = 0$ which is impossible. For the same method, we could conclude that the system must have two symmetry equilibrium points.

To get more information about the equilibrium points, we fix $\lambda = 0.05$ and $h_z = 10$. Test the stability of the equilibrium points for different values of $\alpha$ and $\psi$ by calculating the eigenvalues of Jacobian Matrix at each point.

The result turns out that the eigenvalues are also symmetry, such that if one of the equilibrium points is a stable focus, then the other is an unstable focus. Also, for the spin is restricted in the sphere, any local part of the sphere can be viewed as a plane, so at each point, we only have two non-zero eigenvalues.

We then use the exhaustive method to test large quantity of conditions to find out that they all have one unstable focus and one stable focus. Which means that in general, in this model, no matter what the parameter or initial conditions take, the spin would always turn to the stable focus finally.

In this case, we change the parameter of the external field to be a periodic continuous function like sine or cosine. As show before, this is a really "stable" model, the existence of two equilibrium points does not depend on the parameters. Also, under a large range of choosing of parameters, the system only has one stable focus. So we have the idea that, if we change some of the parameter to be a continuous periodic function, then the stable focus (attractors) would not change so much. Since the trace of the attractor would be a closed curve, then there would also exists a limit cycle around that closed curve.

Firstly, we make a small perturbation to the external field such as $H = (0,0, h_z + \gamma \cdot \sin 2\pi f t)$, where $\gamma$ is a relatively small term. The Fig.1 is an example, showing that the projection of the stable solution (blue closed curve) and the trace of attractor (orange line) on $x$-$y$ plane.

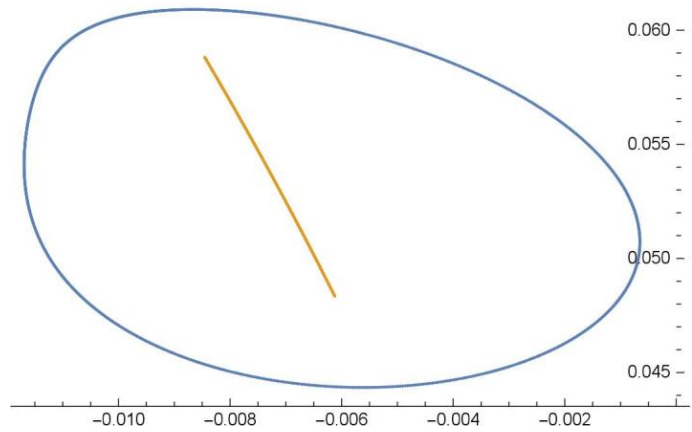

*Figure 1. The projection of the stable solution (blue) and the trace of attractor (orange) on x-y plane, under a small oscillating perturbation of external field.*

Also, we let $H = (0,0, h_z \cdot \sin 2\pi f t)$. The following figure is the projection of the stable solution (Blue) and the trace of attractor (Orange) on $x$-$y$ plane.

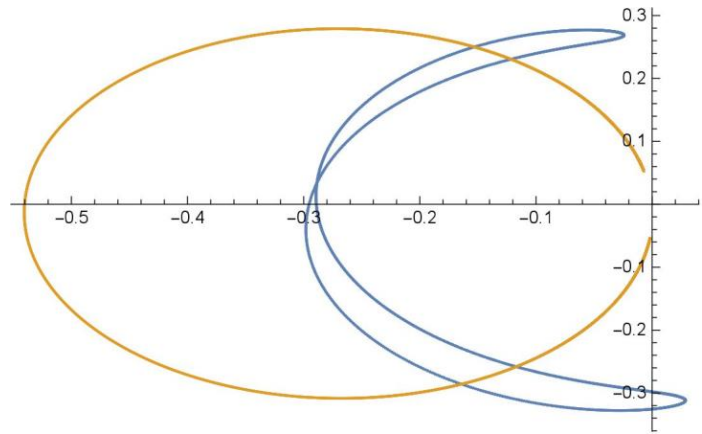

*Figure 2. The projection of the stable solution (blue) and the trace of attractor (orange) on x-y plane, under an oscillating external field.*

Notice that, if we pick up the initial point on the start of the trace of the attractor, we can not get a limit cycle coincide with the trace. For that the "attracting" ability of the attractor is not all the same, and lose control of the spin at some time. The Fig. 3 is an example (the parameters are the same with the previous):

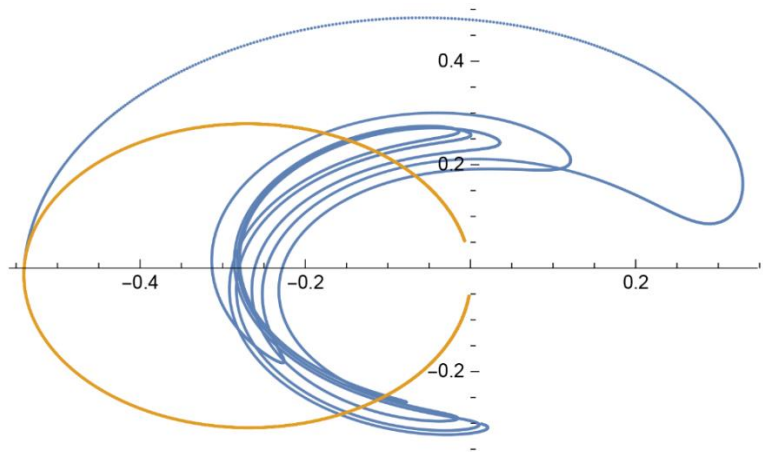

*Figure 3. Sensitive at the initial points on the start of the trace of the attractor. The projection of the solution (blue) and the trace of attractor (orange) on x-y plane, under an oscillating external field.*

Also, we get the relation between the frequency of the external field and that of the spin by stimulating. We calculating the peaks of the solution projecting on $x$-axis, obviously, one peak means one period of the solution. The result in Fig. 4 shows that, the frequency of the external field is always equal to that of the spin.

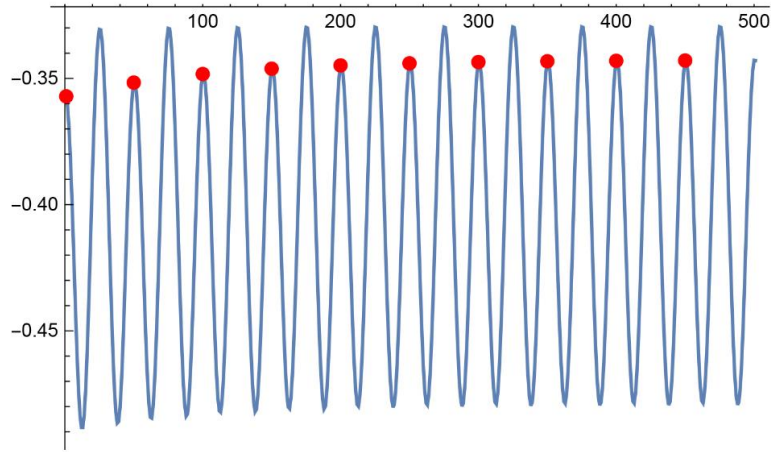

*Figure 4. Period of the spin oscillation is same as the period of external field.*

## 4. Conclusion and remarks

We have studied the evolution of magnetization of a macrospin based on LLG equation with external field and spin transfer torque. In the case of only external field, and constant field plus spin torque, the system always has a stable solution, namely, the spin evolves to a stable focus no matter what the parameters or initial conditions taken. If we extend the model to consider a continuous periodic field, the stable focus (attractors) would not change so much, and always a closed curve and there would also exists a limit cycle around that closed curve except some special points on the attractor. The nonlinearity of the magnetization of macrospin will help understand the dynamic behaviors of nanomagents of domain wall motion under oscillating field and current. Further research needs to be done for lattice and continuum spin systems.

## Acknowledgements

This work was supported by the grants from the National Natural Science Foundation of China (No.11204245), and the Natural Science Foundation of Jiangsu Province (No.BK2012637).